\documentclass[preprint,12pt]{elsarticle}

\usepackage{graphicx}
\usepackage{amssymb}

\journal{arXiv - Version 2.01}

\begin{document}

\begin{frontmatter}


\title{Law on the Market? Abnormal Stock Returns and Supreme Court Decision-Making}




\author[1,2]{Daniel Martin Katz}
\author[1,2]{Michael J. Bommarito II}
\author[3]{Tyler Soellinger}
\author[3]{James Ming Chen}

\address[1]{Illinois Institute of Technology - Chicago Kent College of Law}
\address[2]{CodeX - The Stanford Center for Legal Informatics}
\address[3]{Michigan State University College of Law}

\begin{abstract}
What happens when the Supreme Court of the United States decides a case impacting one or more publicly-traded firms?  While many have observed anecdotal evidence linking decisions or oral arguments to abnormal stock returns, few have rigorously or systematically investigated the behavior of equities around Supreme Court actions.  In this research, we present the first comprehensive, longitudinal study on the topic, spanning over 15 years and hundreds of cases and firms.  Using both intra- and interday data around decisions and oral arguments, we evaluate the frequency and magnitude of statistically-significant abnormal return events after Supreme Court action.  On a per-term basis, we find 5.3 cases and 7.8 stocks that exhibit abnormal returns after decision. In total, across the cases we examined, we find 79 out of the 211 cases (37\%) exhibit an average abnormal return of 4.4\% over a two-session window with an average $|t|$-statistic of $2.9$.  Finally, we observe that abnormal returns following Supreme Court decisions materialize over the span of hours and days, not minutes, yielding strong implications for market efficiency in this context.  While we cannot causally separate substantive legal impact from mere revision of beliefs, we do find strong evidence that there is indeed a ``law on the market'' effect as measured by the frequency of abnormal return events, and that these abnormal returns are not immediately incorporated into prices.
\end{abstract}


\end{frontmatter}


\section{Introduction}
\label{S:1}

On June 13, 2013, the Supreme Court of the United States delivered its opinion in the highly-anticipated bio-patent case of \textit{Association for Molecular Pathology v. Myriad Genetics Inc.}, 133 S. Ct. 2107  (2013).  In this case, the Court considered the important question of whether human genes could be patented.  The defendant in the case, Myriad Genetics, was sued over its patent claims relating to two genes -- BRCA1 and BRCA2 -- whose mutations have been linked to increased risk for breast and ovarian cancer.  Relying on its patent claim, Myriad Genetics had sought to be the exclusive provider of ``BRAC analysis'' and ``BART analysis'' tests used to screen patients for cancer.

\begin{figure}[h]
\centering\includegraphics[width=0.9\linewidth]{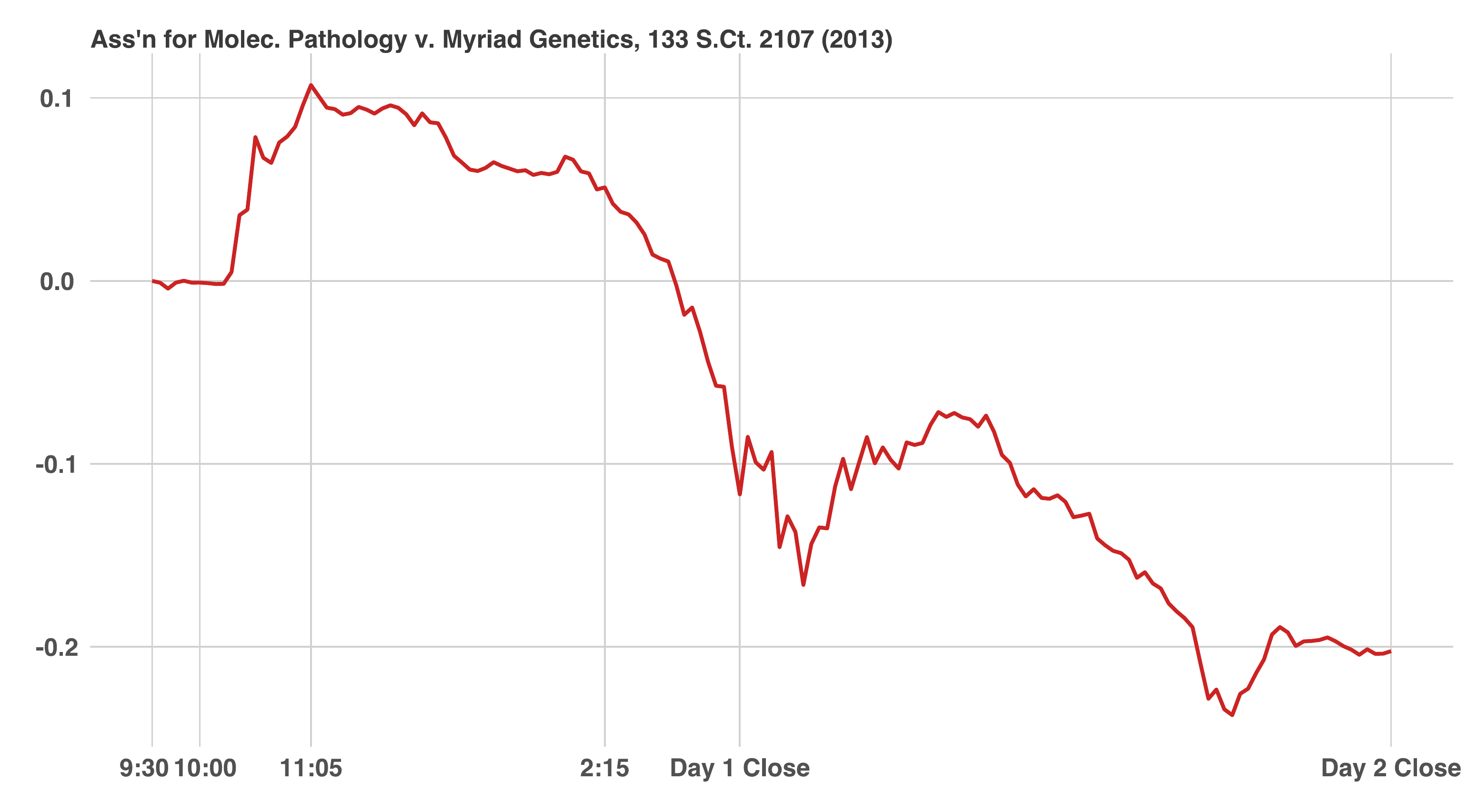}
\caption{Myriad Genetics (MYGN) cumulative abnormal returns from June 13-14, 2013}
\label{myriad_return}
\end{figure}

Ultimately, the Court's decision was seen as a compromise, as it held that DNA sequences fall outside the definition of patentable subject matter under 35 U.S.C. \S 101, but cDNA (complementary DNA) sequences, which do not occur in nature absent human intervention, may indeed be patented.  This compromise was significant not just for patent law broadly, but more specifically for Myriad's business model.  As displayed in Figure \ref{myriad_return}, the market initially ``got it wrong.''  Fueled in part by inaccurate media reports, in the initial hours after the opinion was released, market participants interpreted the Court's decision as positive for Myriad.  In reality, however, the decision was harmful to Myriad as it resulted in the loss of market exclusivity for its BRCA testing revenues.  Eventually, the stock began to trade down in the second half of the session as consensus came around to this understanding.  Contemporaneous media coverage colloquially described the session as a ``wild ride'' and a ``market whipsaw.'' 

As highlighted in Figure \ref{myriad_return}, Myriad's stock fell more than 20\% over the two-day window, a significant move even after controlling for the overall market.  In addition to this significant price change, there was also over a ten-fold increase in daily volume compared to the prior month's average.  For Myriad, the actions of the Supreme Court were clearly related to a material revaluation of their stock.  But how common are events like these, and when they do occur, what is the typical magnitude of corresponding market movement?   While scholars have studied the equity market impacts of individual cases or particular areas of law, none have yet studied a court or market over time and across areas.  

In this paper, we document the frequency and magnitude of abnormal stock returns following Supreme Court action.  We comprehensively review every case before the Supreme Court of the United States, beginning with the October 1999 term and ending with the October 2013 term.  We identify all cases where the parties involved are publicly-traded or the legal question has economic bearing on one or more publicly-traded companies.  

Using this expertly-coded subset of cases, we apply event study methodology to examine returns around the date of decision and oral argument.  For decisions, which are typically released at a known time during market hours, we rely on five-minute intraday OHLC data over a two-day window; for oral arguments, whose transcripts are released same-day but with less regularity of timing, we rely on daily OHLC data over a 60-day window.  While we cannot separate the measurement of substantive legal impact from mere revision of prior beliefs, we can identify cases in which at least one of these effects is present.

On a per-term basis, we find 5.3 cases and 7.8 stocks that exhibit abnormal returns after decision; in total, we find 79 out of 211 cases (37\%) exhibit an average abnormal return of 4.4\% over a two-session window with an average $|t|$-statistic of $2.9$.  In our analysis of oral arguments, we find that the magnitude of abnormal returns around oral argument and decision are strongly rank-correlated, although the returns \textit{per se} exhibit no dependence.  Finally, we observe that abnormal returns related to Supreme Court decisions materialize over the span of hours and days, not minutes, with implications for market efficiency in this context.  We find strong evidence that there is indeed a ``law on the market'' effect as measured by the frequency of abnormal return events, and that these abnormal returns are not immediately incorporated into prices.

\section{Research Overview}
\label{S:2}
Economists have long been interested in how markets respond to and incorporate new information. In finance and law, these questions have typically been addressed using event study methods applied to price data.  Building on the initial work of \cite{bachelier1900theorie}\ and \cite{cootner1964random}, \cite{fama1970efficient} outlines the efficient market hypothesis (EMH). EMH argues that the stock market is informationally efficient, and thus the price of a security reflects available information.  While there are stronger and weaker versions of EMH, semi-strong EMH implies that publicly-announced Supreme Court decisions and oral argument transcripts should be rapidly incorporated into prices.  

Supreme Court decisions and oral arguments may add to the information set related to individual securities, sectors, and markets as a whole.  In some cases, this new information may not be so new, i.e., it may confirm existing market expectations; for example, in cases where the lower court has ruled against a firm and market participants strongly believe the Supreme Court will affirm the prior ruling, then the firm's stock price should, in theory, already reflect the decreased valuation.  If the Court does affirm the lower court, then no significant revision of beliefs should occur, and therefore no significant change in valuation should occur.

In other cases, however, new information may either contradict expectations or collapse a high-entropy state.  In the former example, the Court ``surprises'' and the market must adjust its expectations.  For example, if the lower court has ruled against a firm and market participants strongly believe the Supreme Court will affirm the prior standing, then the firm's stock price should already reflect the decreased valuation.  But if the Supreme Court then unexpectedly reverses the lower court decision in favor of the firm, the price should then increase to remove or ``undo'' the discount created by the lower court decision.  Since beliefs are significantly revised, a proportional change in valuation should therefore occur.

The most complex case involves ``coin-toss'' or ``high-entropy'' cases.  Take, for example, a case in which market participants are perfectly split in their expectations.  One half expect a favorable ruling that will result in a 10\% increase in the stock; the other half expect an unfavorable ruling that will result in a 10\% decrease in the stock.  All else equal, we should then expect the a 10\% abnormal return after decision regardless of whether the decision is favorable or unfavorable.\footnote{In general, the intrinsic uncertainty or predictability of Supreme Court decisions is still a very much an open and unexplored question.  In related work, several of the authors of this study have examined the predictability of Supreme Court decisions using algorithmic approaches \cite{katz2017general}.}
  
We cannot directly observe participant expectations and information sets.  As a result, we cannot systematically disentangle expectation confirmation, surprise, and high uncertainty or entropy cases in our sample.  When we measure abnormal price returns after decision or oral argument,  we may therefore be capturing the direct impact of substantive legal change, the mere \textit{revision of beliefs}, or both.  As documented in \cite{hennessy2015beyond}, this limits our ability to interpret the ``treatment effect'' and underlying causality.  It does not, however, limit our ability to document cases of abnormal return themselves.  

Furthermore, from a practioner's perspective, our price-based analysis is identifying only a subset of events that may present opportunity.  For example, while confirming, surprising, and high-entropy cases may be difficult to identify using only the underlying security price data, option prices and volumes may provide a better glimpse into participant beliefs.  Similarly, while we identify potential inefficiencies in underlying security prices, there are likely many more option-based strategies such as long or short straddles worth investigation.

Treatment isolation aside, event study methods are the typical approach used to explore how markets respond to new information.  The event study methodology pioneered in \cite{fama1969adjustment} and further outlined in \cite{brown1985using}, \cite{campbell1997econometrics}, \cite{mackinlay1997event} and \cite{bhagat2002event} is designed to test for the presence of abnormal returns associated after an event or change in information set.  While important work such as \cite{hennessy2015beyond} has recently cautioned against inappropriate interpretation of event studies,  they have become a staple of economics and finance, with thousands of examples across a wide range of data.  Classic applications include the exploration of market reactions to information updating events including announcements of earnings \cite{firth1976impact}, \cite{patell1981ex}, dividends \cite{kalay1985predictable}, \cite{kalay1986informational}, stock splits \cite{charest1978split}, \cite{lamoureux1987market}, mergers \cite{asquith1983merger} and tender offers \cite{dodd1977tender}.   

Historically, analysis of abnormal return events has also included the exploration of third-party or non-market actors or events, such as market reactions to decisions of the Federal Reserve \cite{bomfim2003pre}, \cite{bernanke2005explains}, administrative agencies \cite{bosch1994wealth}, \cite{sarkar2006market}, \cite{lax2006courts}, \cite{ellert1976mergers} legislatures, \cite{gilligan1988complex}, \cite{cutler1988tax}, and elections \cite{roberts1990political}, \cite{herron1999measurement}, \cite{knight2006policy}.  Within the narrow context of litigation and judicial decision-making, a number of event studies have been carried out in corporate law \cite{bhagat1998shareholder} \cite{ryngaert1988effect}, tax law \cite{key2011shareholder} \cite{dhaliwal1998wealth}, patent law \cite{marco2013certain}, environmental law \cite{sun2011effects}, communications law \cite{chen2013does},  products liability \cite{prince2002effects} \cite{viscusi1990market}, antitrust \cite{huth1989impact} \cite{bizjak1995effect} and property rights \cite{keay2011property}.  

Collectively, these studies offer evidence that the decisions of legal institutions and actors impact the securities markets.  However, there are some serious limitations in these prior papers, including limited longitudinal scale and breadth.  Our primary purpose in this research is to address these two dimensions by evaluating every case decided by the most visible court in the United States over a fifteen year period.

\section{Data}
\label{S:3}
\subsection{Supreme Court Data}
Although the precise date of decision is often not known in advance, Supreme Court decisions, much like earnings reports, are anticipated events that may reveal new information.  Unlike earnings events, however, most finance professionals are not familiar with the process that produces these decisions.

Supreme Court cases typically involve an appeal from a prior decision.  In the vast majority of cases, one or more lower courts have already heard claims from the parties involved in a dispute and have issued public decisions.  Generally, the Supreme Court may take cases on appeal for a number of reasons, including appeal to reverse a lower court decision which in some cases may help resolve inconsistency between federal courts.  Since such disputes are already public and typically originate from a prior decision, market participants can readily form beliefs and adjust valuations if warranted.  Furthermore, as required by SEC regulations and the Securities and Securities Exchange Acts, companies are required to disclose material events, including pending litigation, in 8-K and annual 10-K reports.

Appeals from lower courts are not automatic or guaranteed, however.  A party may petition the Supreme Court to hear a case; the Court may then calendar the petition for review in conference.  Conferences are held periodically, and, based on the norms of the Court, if a minimum number of Justices desire to hear the case, the petition may be granted.  At this point, the case may be docketed for oral argument.  Throughout this period, both the parties themselves and third-parties, such as business associations or interest groups, may file briefs arguing for or against the parties.  Eventually, oral arguments are heard, and, in some cases, the Justices may request additional time to allow the parties to re-argue.

From this point on, market participants anticipate that the Court may issue an opinion on any designated Monday, Tuesday, or Wednesday between 10:00-10:15AM Eastern.\footnote{Given the days of week on which decisions are announced, almost none of our two-session samples span weekends.}  Realistically, even in the simplest 9-0 majority cases, the preparation of opinions may take at least a few weeks.  In cases where there are multiple concurring or dissenting opinions, this timeline may extend to months.  Court-watchers therefore know that, once oral arguments have been heard, they must monitor the Supreme Court press coverage each day calendared for decisions.  These release windows are always during US open market hours.

As we define it, ``law on the market'' (LOTM) exists when the action of a legal institution effects an abnormal return in the price of one or more publicly-traded companies. In order to identify LOTM events, we reviewed Supreme Court cases and first classified them into two categories: LOTM Candidates and Non-Candidates.  Much of the Supreme Court's docket is filled by cases with little or no direct market relevance. For example, as a general matter, cases involving criminal procedure, capital punishment cases, or jury composition impose no obvious discernible economic impact on publicly-traded firms.  Our expert coders reviewed each case, only designating the case as a LOTM candidate case if the parties or legal ramifications could plausibly be linked to relevant market participants.  

Our prompts took the form of two questions.  First, could this decision of the Supreme Court of the United States plausibly affect publicly-traded companies or sectors?  If so, then which publicly-traded companies or sectors?  We recognize that our event definition and coding do not capture all potential economic impacts from the Court's actions.  For example, an affected party might be held privately, might be a non-profit institution, or the economic consequences might only be understood or apparent over a longer period.  In these instances, we would not be able to identify abnormal returns after Court actions. Also, despite our best efforts to maximize recall and include any tangentially-related cases, we acknowledge that we may have failed to identify some candidates from the 1,300+ cases we reviewed.  Thankfully, these shortcomings imply that our overall results represent merely a \textit{lower} bound, not an \textit{upper} bound, on the frequency of ``law on the market'' events.  In total, among the 1,363 total cases reviewed in our sample, we identified 211 candidate LOTM cases plausibly affecting one or more firms or sectors.

\subsection{Market Data}
Each of our 211 candidate cases is associated with one or more firms or sectors.  For firms, we identify the symbol and exchange at the time of event; for sectors, we select exchange-traded funds (ETF) from the Select Sector SPDR family, as they have the longest history and most liquid trading.  We then collect market data around both decision and oral argument for each case and each symbol.  For decision, we retrieve OHLC data at five-minute intervals for the period one week prior to and one week after the decision.  This provides us with an ample pre-event and post-event estimation window.  For oral arguments, we retrieve OHLC data at daily intervals for the period one month prior to argument and one month after.  In addition, we retrieve price and volume data for the S\&P 500 exchange-traded fund \textsc{SPY} for both five-minute and daily periods matching each case sample.

It is worth noting that most event studies are performed on just \textit{inter}day, not \textit{intra}day, data.  Scholars have typically conducted event study analysis using daily, weekly, or monthly price data. Nearly every event study in law, and the vast majority of analysis in finance, leverages data with at most daily frequency.  Interday studies, however, face a difficult and unavoidable tradeoff: either a researcher must rely upon relatively few data points, or they must collect data over longer durations in order to generate a sufficient statistical sample. As the estimation windows are extended for sample size, it becomes more difficult to defend the link between event and measurement.  With every passing day, the chance of confounding events grows.  Whether it is a subsequent earnings announcement, a change in senior management, updates in the status of a regulatory approval, a merger announcement, or some other important change, the potential for misidentification and measurement error looms large.  As a result, interday studies must attempt to manually identify confounding factors and attempt to control for them, only increasing model complexity.

At its core, we believe the problem is one of data, not one of methods. Scholars in the literature increasingly agree.  As noted in \cite{kothari2007econometrics}, ``short-horizon methods are quite reliable and while long-horizon methods have improved, serious limitations remain.'' We likewise believe that intraday data is the key to avoiding Type I and II error in the estimation of abnormal returns around an event with narrow temporal scope.  Given that access to granular price data, including minutely and tick data, has become more affordable, we have collected our data for decision events based on this premise.

\section{Methodology and Results}
\label{S:4}
\subsection{Abnormal Returns}
In order to test for abnormal returns, we must first select a model of \textit{normal} returns.  While there are many models for asset pricing and stock returns,\footnote{See \cite{nadarajah2012models} for a review of candidates.}  our model must satisfy three constraints: (a) stable estimation for samples with $N \leq 175$, as dictated by our five-minute bar data and windows; (b) simple comparison and communication of results, including degree of uncertainty or belief; and (c) application across a wide range of firm sizes, industries, and time periods.  Constraints (a) and (b) preclude models requiring higher-order moments or more sensitive, asymmetric distributions, such as skewness, kurtosis, or the Variance-Gamma family.  Constraint (c) also counsels against tools like multi-factor models, which require firm size, industry, and period-specific factors and calibration.

Given these constraints, we select the most common approach in finance - the Capital Asset Pricing Model (CAPM).  Since our analysis is short-term, focusing on intra-day timescales, we also take the risk-free rate to be constant.  Critically, throughout the entirety of our study sample, the daily overnight rate has not exceeded (0.02\%).  The end result is that CAPM, in our case, collapses to the simple market model of returns where $r_f=0$.

Estimating the impact of an event under the market model is detailed in many papers cited above, and \cite{mackinlay1997event} is the most widely-cited across event study literature; the interested reader is directed there for a conceptual introduction to our methodology.  To perform this procedure, we rely on Sun's implementation, \textsc{evReturn} in the R package \textsc{erer}; this package was developed in \cite{sun2011effects} based upon approaches previously outlined in work such as \cite{fama1969adjustment} and \cite{mackinlay1997event}.\footnote{Code necessary to replicate this model is available on Github at https://github.com/mjbommar/law-on-the-market. However, due to terms of service associated with some of the stock data, we are not able to make intraday stock data publicly available.} 

\begin{figure}[h]
\centering
\includegraphics[width=0.9\linewidth]{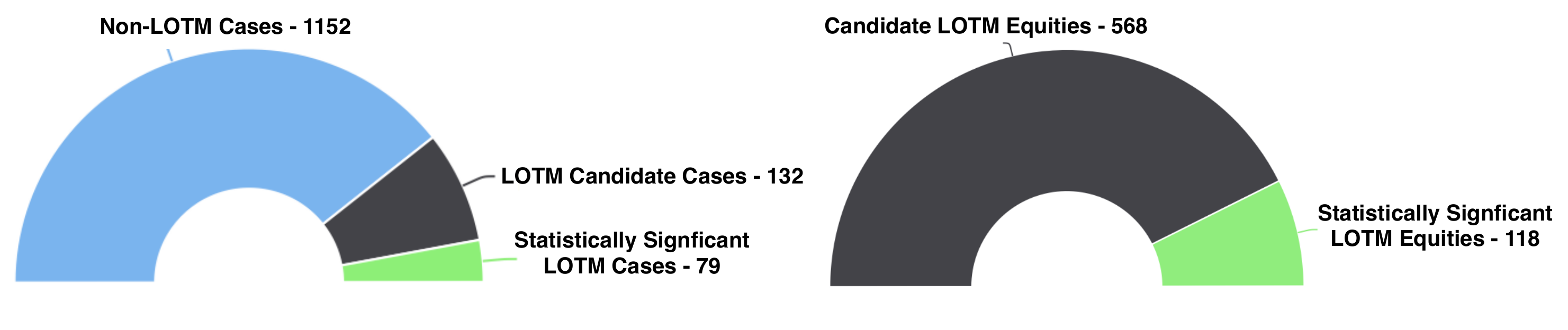}
\caption{Distribution of LOTM and Non-LOTM Cases and Securities (1999-2014)}
\label{lotm_candidate_distribution}
\end{figure}

Applying this estimation framework to our 211 candidate LOTM cases, we find that 79 cases (37\%) exhibit an average abnormal return of 4.4\% over a two-session window with an average $|t|$-statistic of $2.9$.\footnote{There are at least twenty stocks that are significant against one or more ETFs but that are not significant against the S\&P.  All tables and figures are calculated using only ticker symbols for securities or exchange traded funds which are statistically significant with respect to the S\&P500 index.} Figure \ref{lotm_candidate_distribution} summarizes the total number of cases and securities, number of candidate cases and securities, and number of significant cases and securities.\footnote{See the Github repository at https://github.com/mjbommar/law-on-the-market for a complete \textit{Table of Cases} listing every case for which we detect abnormal returns.}  Treating our tests of these 211 cases as independent, the probability of observing 79 significant cases at a $p$-value threshold of $0.05$ by chance is exceedingly small; we can see this by modeling our aggregate outcome as a binomial distribution with $X \sim B(N=211, p=0.05)$ and evaluating $\mathbb{P}(X < 79)$, which is greater than $1 - 10^{-40}$\footnote{See, e.g., BinomialCDF[211, 0.05, 79] on Wolfram Alpha.}.  At the stock-level, we performed 1573 tests with 298 positives; $\mathbb{P}(X < 298)$ is similarly $\approx 1$ at $> 1 - 10^{-80}$.

As displayed in Figure 3 below, over our fifteen-year sample, the frequency of both potential and realized LOTM events has been fairly steady.  On an annual basis, there were an average of 5.3 LOTM cases per term and 7.8 statistically significant LOTM securities.  While there is some annual variation in the number of candidate and significant events, Figure 3 highlights that most years are fairly close to the average with the October 2009 term featuring only 1 LOTM case but with the immediately following year of 2010 serving as the largest year with 10 LOTM cases.  
\begin{figure}[h]
\centering
\includegraphics[width=0.9\linewidth]{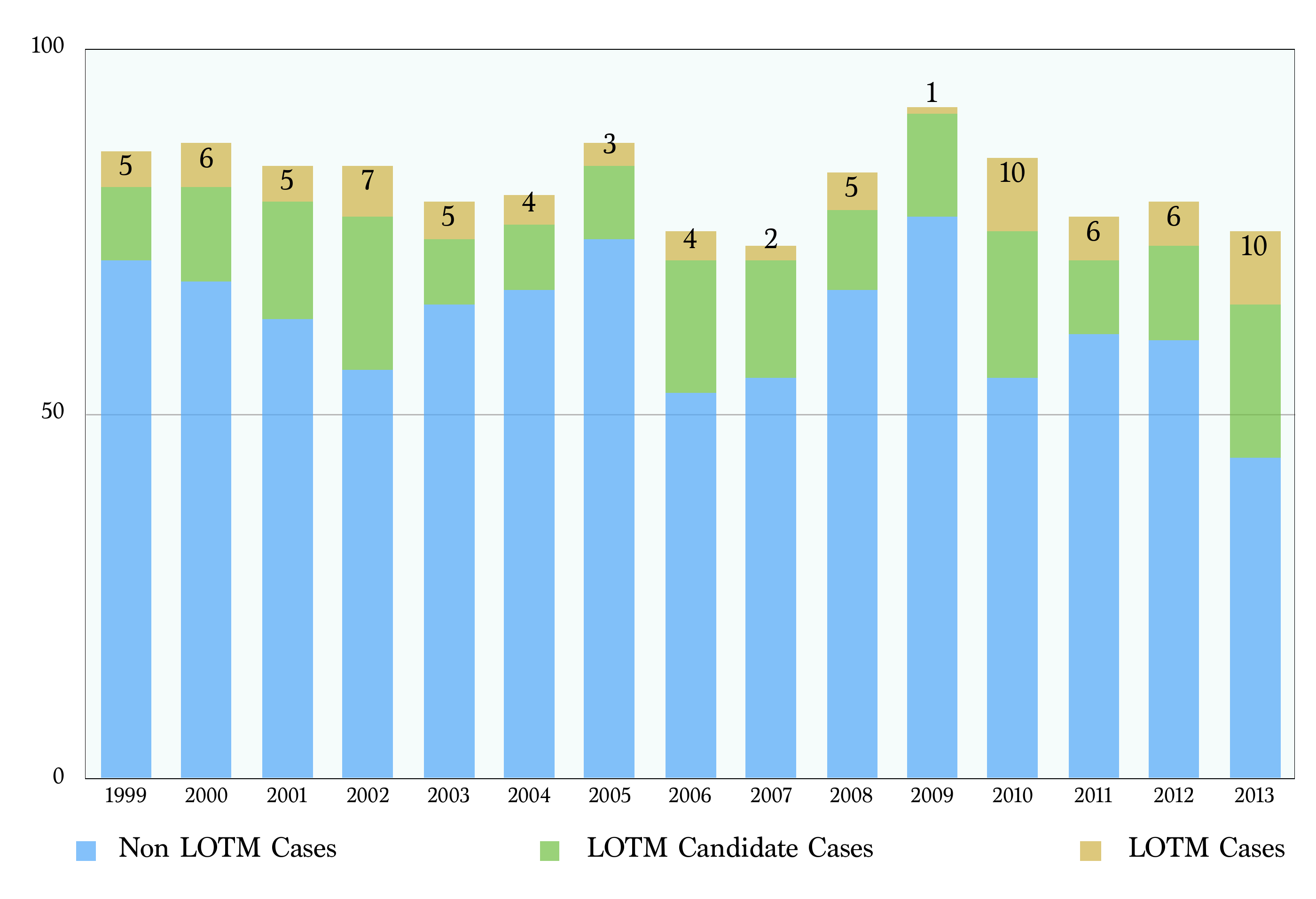}
\caption{Annual Frequency of LOTM Cases (1999-2014)}
\label{lotm_event_frequency}
\end{figure}

\subsection{Dollar Impact}
Returns are not a measure of dollars or wealth, however, and while an individual or institution may or may not sell the security to realize the gain or loss, the mark-to-market wealth of individuals and entities does change.  In this study, we observe a wide variety of companies with relevant issues before the Court. These companies vary especially with respect to market capitalization, and this range of capitalization impacts the size of the observed returns.  Importantly, these variations may mean that a 1\% abnormal return may have very different dollar or wealth impacts on market participants.  Theory suggests that larger, more diversified companies would be less affected by decisions of equal economic impact than smaller, less diversified companies.  For example, a Court decision affecting revenues of one business unit should have less impact on a firm with many business units than one with few, on average.  

In order to evaluate these dollar impacts, we calculate the change in equity value for each of our statistically significant abnormal return events.  After accounting for changes in number of shares outstanding, this is measured by the difference between market cap at the beginning and end of our event windows.  Using this approach, we estimate the change in total wealth for each statistically significant event. In aggregate, \textit{law on the market} events exhibit total wealth changes in excess of 148 billion dollars in our sample.  On average, this is 9.91 billion dollars per year, or 1.88 billion per event.  Figure 4 offers the distribution of these wealth changes as a function of time.  Clearly, these events are significant in real dollar terms, not just percentage returns.

From a dollar impact perspective, the largest event is \textit{United States v. Locke}, 120 S. Ct. 1135 (2000), a supremacy clause case from the October 1999 term.  In this case, the Court held that ``regulations regarding general navigation watch procedures, crew English language skills and training, and maritime casualty reporting are preempted by the comprehensive federal regulatory scheme governing oil tankers." Among other things, this case was significant as it limited a set of state oil transport regulations passed in the wake of the Exxon Valdez oil spill. After the decision, which relieved oil transporters from complying with certain state-level regulations,  the value of Exxon Mobil Corporation (XOM) increased by more than 23 billion dollars and the value of Chevron (CVX) by more than 3 billion dollars.

\begin{figure}[h!]
\centering
\includegraphics[width=0.9\linewidth]{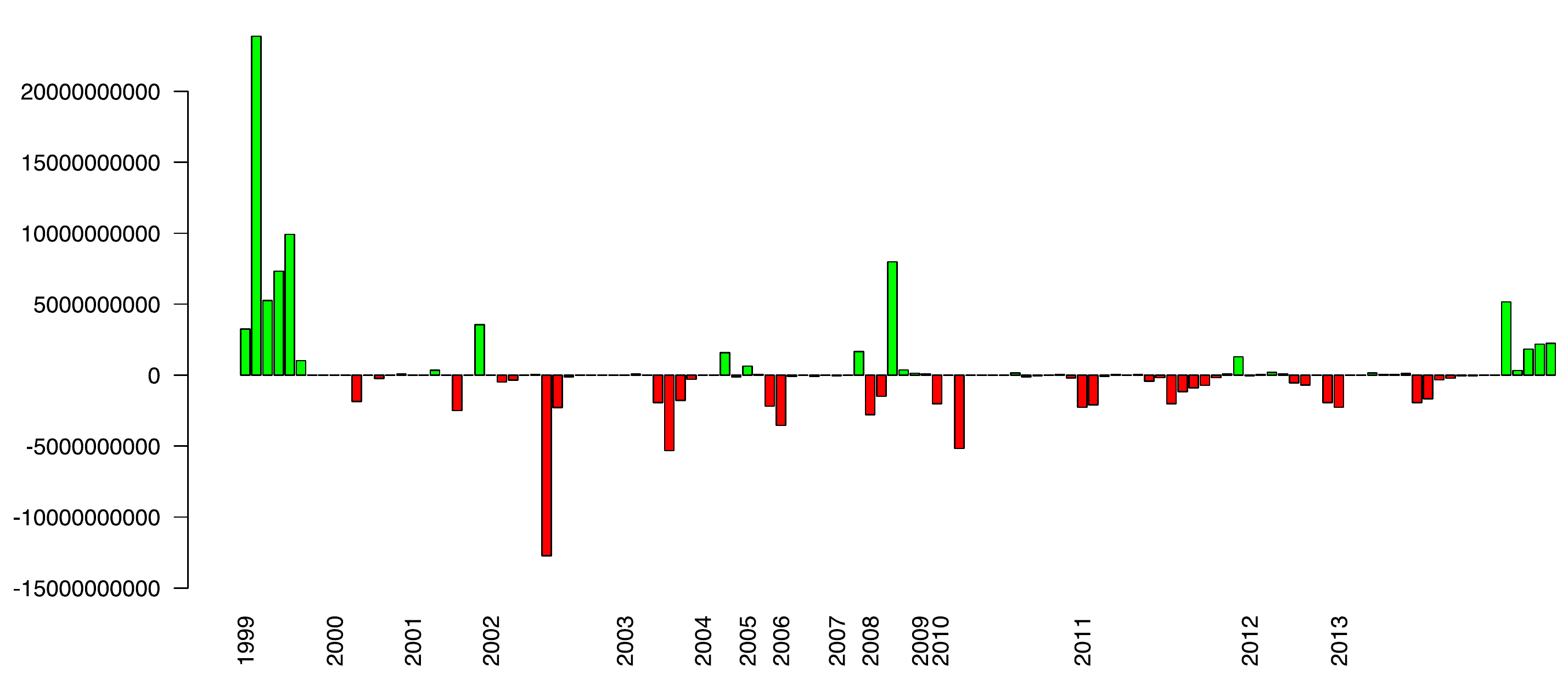}
\caption{Distribution of Wealth Effects as a Function of Time}
\label{lotm_event_frequency}
\end{figure}

\subsection{Oral Arguments and Return Factors}
We have clearly documented the presence of abnormal return events after decisions and measured their dollar impact on wealth.  But what drives these abnormal returns?  Are they just a function of revision of beliefs alone? Are there any factors that can foretell these events or their magnitude?

We investigate these questions through two analyses.  First, we assess excess returns around oral argument, investigating how these correlate with subsequent post-decision returns.  Second, we investigate how factors such as the area of law or voting patterns impact the presence or magnitude of abnormal returns. Together, these analyses can provide additional insight to the case for or against law on the market.

Prior to the date of decision, there are many opportunities for market participants to incorporate information about judicial activity. These include lower court decisions, grant of \textit{certiorari}, filing of \textit{amicus} briefs, and oral argument or re-argument.  In particular, oral arguments typically receive the most attention from the press and Court observers.  While oral argument transcripts have been released same-day for almost all arguments, we have no way to ``timestamp'' their actual release or general availability to the public intraday.  As a result, we analyze returns around oral argument at daily frequency, not intraday, and the results are not directly comparable to our post-decision returns.

In total, we identify 1,168 samples with price data available around oral argument.  We calculate excess return relative to the S\&P 500 for (a) the day before argument to the day after argument, (b) the week before argument to the week after argument, and (c) the month before argument to the month after argument.  Then, for each of these symmetric daily, weekly, and monthly windows, we examine the Pearson and Spearman correlation between excess returns around oral argument and abnormal return post-decision.

As the results in Table \ref{oral_argument_correlation} demonstrate, most pairwise relationships are not strongly dependent.  For example, all pairwise correlations against the post-decision abnormal return $D$ \textit{per se} are below 15\%.  Looking, however, at the \textit{absolute value} of $D$, i.e., the magnitude of post-decision abnormal return, paints a different picture.  In particular, the relationship between the magnitudes of oral argument return $|A|$ and post-decision return $|D|$ is much stronger.  Both the Pearson correlation $\rho(|A|,|D|)$ and Spearman correlation $\rho_s(|A|,|D|)$ are between 20\%-40\% for all oral argument windows.  

\begin{table}[h!]
\centering
\begin{tabular}{|r|r|r|r|}
\hline
{} &  Day & Week & Month\\\hline
$\rho(A,D)$     &             0.081724 &             -0.043022 &               0.060645 \\
$\rho_s(A,D)$     &             0.031031 &             -0.015607 &               0.043728 \\
\hline
$\rho(A,|D|)$ &             0.233393 &              0.181999 &               0.180640 \\
$\rho_s(A,|D|)$ &             0.085366 &              0.071351 &               0.091160 \\
\hline
$\rho(|A|,D)$     &                 0.122116 &                  0.149385 &                   0.056583 \\
$\rho_s(|A|,D)$     &                 0.061151 &                  0.116090 &                   0.051309 \\
\hline
$\rho(|A|,|D|)$ &                 0.386834 &                  0.403502 &                   0.339776 \\
$\rho_s(|A|,|D|)$ &                 0.288358 &                  0.192646 &                   0.240447 \\
\hline
\end{tabular}
\caption{Correlation between excess return around oral argument windows $A$ and post-decision abnormal return $D$.  $\rho$ is the Pearson product-moment coefficient and $\rho_s$ is the Spearman correlation coefficient.} 
\label{oral_argument_correlation}
\end{table}

These correlations imply that in a number of instances market participants incorporate information about the magnitude, but not necessarily the direction, of the Court's decision.  From a coarser perspective, we also simplified the analysis to a two-by-two matrix capturing up-or-down for oral argument and decision in table \ref{oral_argument_crosstab}.  While approximately two-thirds of excess returns are negative around oral argument windows, the direction of change at oral argument does not provide significant information for decision; the direction of post-decision return is still a coin flip.

\begin{table}[h!]
\centering
\begin{tabular}{lrr}
\hline
	& $D < 0$ & $D \geq 0$\\
\hline
$A < 0$ &  0.348630 &  0.303378 \\
$A \geq 0$	&  0.171447 &  0.176546 \\
\hline
\end{tabular}
\caption{Cross-tabulated sign of two-day oral argument and post-decision returns.}
\label{oral_argument_crosstab}
\end{table}

One theory is that oral arguments begin to reveal the preferences of individual Justices sitting on the Court.  If so, then we should see a relationship between the subsequent voting coalition and the direction or magnitude of abnormal return.  For example, in cases where market participants the Court issues a unanimous $9-0$ decision, there was likely less uncertainty about the outcome prior to announcement.  But in cases where the Court rules $5-4$, then the case is likely higher-entropy.  Since the entire decision rests on the decision of one ``swing'' vote, the outcome is often much more uncertain.

To evaluate this theory, we merged all abnormal return results with data on the vote margin and duration between argument and decision, which is a proxy to potential dissent or uncertainty.  We first calculated the Pearson and Spearman correlation coefficients between (a) the estimate and vote margin and (b) the absolute value of estimate and vote margin.  For (a), we observed Pearson and Spearman coefficients of -0.03 and 0.01 with p-values $>$ 0.2 for both measures.  For (b), we observed Pearson and Spearman coefficients of -0.04 and -0.06 with p-values of 0.2 and 0.02, respectively.   Next, we analyzed the mean, median, and standard deviation of estimates conditioned by vote margin; with the exception of cases with zero vote margin, an exceedingly rare case, no differences were observed between other vote configurations.  Given the very small sample size of these ``tied'' vote events ($<$1\%), we cannot read much into this result. 

Next, we merged the abnormal return results with information on the relevant SPDR Sector index (XLU, XLV, XLI, XLF, XLY, XLE, XLB, XLP, XLK) and the SCDB petitioner/respondent codes.  While not all businesses are ``confined'' to a single sector, the SPDR Sector indices represent an accepted market-based definition.  The SCDB petitioner and respondent codes, while based arbitrarily on historical Supreme Court data and researcher choice, are much more fine-grained.   First, we evaluated the means, medians, standard deviations, and IQRs of events based on the SPDR sector index.  We find no statistically significant differences between sectors for the estimate (including direction); Evaluating the absolute value of estimate, i.e., estimate magnitude, we also find no significant differences.  In summary, within our sample's standard errors, abnormal returns do not vary significantly by sector.

We then evaluated the means, medians, standard deviations, and IQRs of events conditioned by petitioner and respondent coding.  Here, we find many more interesting and potentially significant relationships despite the smaller sample sizes by petitioner or respondent code.  For example, some petitioners have significantly left-skewed or negative distributions for abnormal return, including the Department/Secretary of Agriculture, trucking companies, brokers or stock exchanges, and cable TV providers.  Conversely, a number of petitioners have a significantly right-skewed or positive abnormal return distribution, including environmental organizations, sellers or vendors, or the FDA.  While sample sizes within our sample alone are too small for statistical significance, these results corroborate extant research on the topic and justify further research across lower court rulings.

Above, we provide substantial evidence for the existence of ``law on the market'' events, measuring both their frequency and magnitude.  We also investigate returns around the most salient information event prior to decision - oral argument - and how these returns correlate with subsequent post-decision returns.  While we cannot measure the intrinsic uncertainty of the Court's decision as a baseline, we do find a surprisingly low correlation between the direction of oral argument and decision returns.  Market participants seem to understand the magnitude or importance of cases, but do no better than a coin flip when it comes to the actual direction of return post-decision.  This observation does not align with the accuracy rates exhibited by experts in \cite{ruger2004supreme} or algorithms in \cite{katz2017general}, \cite{ruger2004supreme}.

\subsection{Information Incorporation and Market Efficiency}

To further investigate market efficiency, we turn our attention to the \textit{timing} of abnormal return.  In almost all instances, the Court's decision is announced between 10:00-10:15AM ET.  Does the market ``digest'' the opinion by 10:30AM, 11:00AM, or even end of day?  We approach this problem by evaluating the abnormal return time series within our five-minute OHLC bar data, calculating the percentage of total abnormal return that is realized over time.  In the absence of other new firm-specific information, perfect market efficiency would suggest that this time series should jump to 100\% in the seconds or minutes after the Court's opinion is released.  Recalling our motivating Myriad example, however, cautions against this assumption.  But is the Myriad case anomalous or par for the course?  

\begin{figure}[h!]
\centering
\includegraphics[scale=0.4]{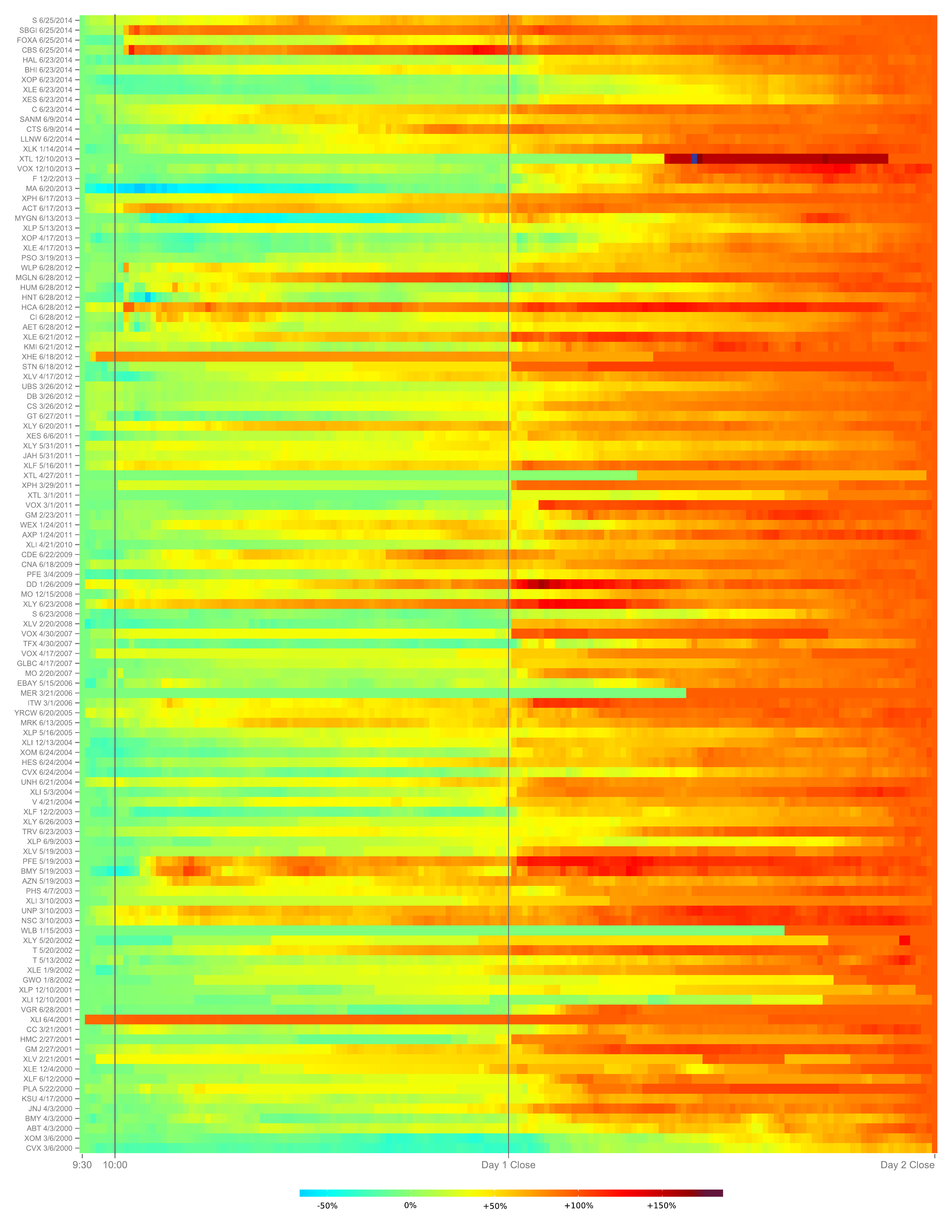}
\caption{Speed of Incorporation of Cumulative Abnormal Returns over Event Window}
\label{abnormal_return_event_window}
\end{figure}

Figure \ref{abnormal_return_event_window} offers a portrait of signal incorporation across all significant LOTM events in our sample.  Each statistically significant security-event pair is displayed on the rows in the y-axis, while the columns along the x-axis correspond to each five minute interval over the two-day window.  The color of each cell corresponds to the percent of total abnormal return that has been realized \textit{up to} that five-minute interval.  We denote key times, e.g., the market opening, approximate 10:00AM ET release, the market close, etc. along the bottom of the x-axis. 

Figure \ref{abnormal_return_event_window} demonstrates substantial variation in the rate at which market participants respond to Supreme Court decisions.  One striking feature, however, is how little abnormal return is observed on the first day of trading.  In many cases, the abnormal return primarily materializes ``overnight'', i.e., between the first day's close and second day's open.  Given the complexity of Court opinions and the timing of thorough press analysis, this may not be so surprising.  Even in incredibly salient and public cases such as the ``Obamacare'' case (\textit{NFIB v. Sibelius})  and \textit{Myriad}, we observe initially incorrect or delayed price incorporation.

Market participants employing high-frequency and algorithmic trading have focused on information events such as central bank announcements  \cite{bernanke2005explains}, \cite{jansen2007were}, consumer sentiment data release \cite{akhtar2012stock}, and other common news announcements \cite{gross2011machines} \cite{schumaker2009quantitative}. In extreme examples, the window for information incorporation has been reduced to mere milliseconds \cite{scholtus2014speed}.  The actions of Courts, however, do not reveal the behavioral signature of an informationally efficient market with rapid signal incorporation.  

\section{Case Studies}
\label{S:5}
\begin{figure}[h!]
\centering
\includegraphics[scale=0.5]{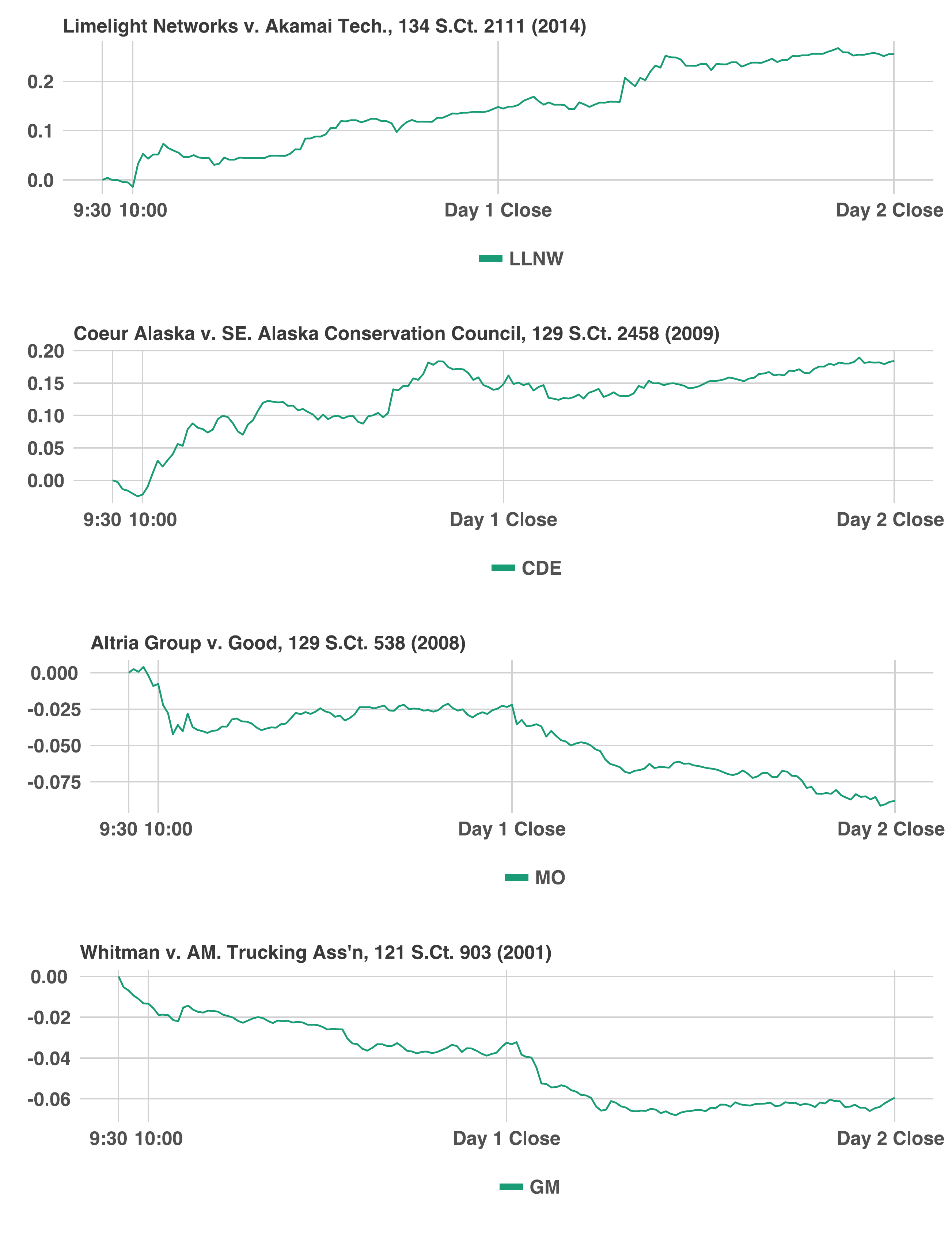}
\caption{Cumulative Abnormal Returns over Event Window}
\label{abnormal_return_event_case_study}
\end{figure}

Each case has a story and it would be impossible to do complete justice to all of the events in our sample.  However, in order to provide the reader with more context, we describe four significant cases which collectively involve questions tort law, preemption, environmental law, administrative law, and patent law. In Figure \ref{abnormal_return_event_case_study}, we display the two-day time series of cumulative abnormal returns, starting at the 9:30AM ET open on the day of decision and concluding at 4:00PM on the day following the Court's decision.

The first pane in Figure \ref{abnormal_return_event_case_study}, \textit{Limelight Networks v. Akamai Tech}, 134 S.Ct. 2111 (2014), represented the largest two-day cumulative abnormal return Limelight's stock posting a cumulative abnormal return of over 25\%.  The patent infringement case considered whether, under the specific circumstances present in the case, an infringement claim could be sustained even if no direct party had committed patent infringement.  The Court held that the answer was no, essentially ``saving'' Limelight Networks (LLNW) from potential bankruptcy. The stock experienced  an almost immediate abnormal return of 5\%, finishing the first session up nearly 15\%.  By the end of the second day, the stock had logged over an additional 10\% abnormal return.

The second pane in Figure \ref{abnormal_return_event_case_study},  \textit{Coeur Alaska, Inc. v. Southeast Alaska Conservation Council}, 129 S.Ct. 2458 (2009), represented a substantial gain for Coeur Mining, Inc. (CDE).  In \textit{Coeur Alaska}, the Supreme Court considered a challenge brought by a series of environmental groups who sought to block the disposal of tailings from a former gold mine.  Coeur Alaska had obtained a permit to dispose of 4.5 million tons of tailings in a local lake located inside a national park. Environmental organizations challenged the permit, arguing it violated the Clean Water Act.  The Supreme Court rejected the challenge, upholding the permit and thereby enriching its parent company.  Within hours, the Coeur stock (CDE) traded up 10\% and finished the day by posting cumulative abnormal returns around 15\%.  The following day saw some small additional gains, but most of the returns had been established in the first trading day.

The third pane in Figure \ref{abnormal_return_event_case_study} shows the abnormal return related to \textit{Altria Group, Inc. v. Good}, 129 S. Ct. 538 (2008), a case originating when ``smokers of `light' cigarettes filed suit, alleging that cigarette manufacturers violated the Maine Unfair Trade Practices Act (MUTPA) by fraudulently advertising that their `light' cigarettes delivered less tar and nicotine than regular brands." The Supreme Court considered whether Maine's statute allowing plaintiffs to pursue a fraud claim was preempted by federal law.  The Court ultimately held that ``a state law prohibiting deceptive tobacco advertising was \textit{not} preempted by a federal law regulating cigarette advertising," thus allowing the state tort claim to proceed.  Following this Court announcement, Altria's (MO) stock immediately trended down, followed by lateral trading for much of the date of decision.  In the following day, Altria's stock continued to decline relative to the market as the financial implications of the Court's decision began to become more widely understood.  Ultimately, at the close of the two-day window, the stock had experienced a negative cumulative abnormal return of nearly 9\%.   

In the fourth pane in Figure \ref{abnormal_return_event_case_study}, \textit{Whitman v. American Trucking Ass'ns, Inc.}, 121 S.Ct. 903 (2001), the Court considered a challenge to the Environmental Protection Agency's National Ambient Air Quality Standard (NAAQS). Among other things, the NAAQS is responsible for regulating permissible amounts of ozone and particulate matter.  In response to revision in national ambient air quality standards, a series of organizations, including the American Trucking Association, challenged the EPA, arguing the enabling statute had impermissibly delegated legislative power to the EPA.  While the Court's decision was mixed with respect to substantive law impact, the market's reaction was clearly negative.  General Motors (GM), at the time, the largest automaker, and Honda Motors (HMC), also traded down over the next two sessions.

\section{Conclusion}
\label{S:6}
In this paper, we document the frequency and magnitude of abnormal stock returns following Supreme Court action.  We comprehensively review every case before the Supreme Court of the United States from the 1999 term through the 2013 term, identifying the subset of cases where the parties involved are publicly-traded or the legal question has economic bearing on one or more publicly-traded companies.  Using this expertly-coded subset of cases, we examine abnormal stock returns around the date of decision and oral argument.  We find that ``law on the market'' events are persistent over time, relatively frequent, and often large in both percentage return and dollar terms.  While we cannot separate the measurement of substantive legal impact from mere revision of prior beliefs, we can identify cases in which at least one of these effects is present.  Furthermore, in our analysis of returns around oral argument and the rate at which abnormal returns are realized, we find evidence of market inefficiency relative to other common information events.

While this paper provides the first broad, longitudinal study of markets and courts, there is ample room for improvement.  Future research would extend this analysis to both higher and lower frequency price data.  At higher frequency, an evaluation of order book and tick data around decision announcement will provide substantially more evidence towards information efficiency and market attention to these events.  At lower frequency, a more holistic view of disputes, beginning with the original filing in state or Federal court, can provide more information about how market participants perceive court activity and revise their beliefs as new information becomes available.  We can also incorporate option price and volume data, where available, to better understand participant beliefs and the implied case uncertainty.  As researchers increasingly identify law and finance as an interconnected system-of-systems \cite{ruhl2017harnessing}, we look forward to more research and attention to ``law on the market'' phenomena.

In conclusion, this paper has sketched a first portrait of the \textit{law on the market} phenomena. Evaluating the population of cases decided in recent terms of the Supreme Court of the United States, we identify a substantial number of cases where the price of one or more legally implicated, publicly-traded companies exhibit significant abnormal returns after Supreme Court decision.  Furthermore, these abnormal returns are realized on timescales that suggest delays in information incorporation.  We look forward to additional interest and research in event-driven arbitrage opportunities and increasing market efficiency in this space.
\\

%





\bibliographystyle{model1-num-names}
\bibliography{lotm.bib}







\end{document}